\documentclass[twocolumn,aps,pra,showpacs,superscriptaddress]{revtex4}
\usepackage{graphicx}
\usepackage{amsfonts}
\usepackage{amsmath}
\usepackage{amssymb}
\usepackage{color}

\begin{document}

\title{Proposed experiments of qutrit state-independent contextuality and two-qutrit contextuality-based nonlocality}



\author{Ad\'an Cabello}
 \affiliation{Departamento de F\'{\i}sica Aplicada II, Universidad de
 Sevilla, E-41012 Sevilla, Spain}
 \affiliation{Department of Physics, Stockholm University, S-10691
 Stockholm, Sweden}
\author{Elias Amselem}
 \affiliation{Department of Physics, Stockholm University, S-10691
 Stockholm, Sweden}
\author{Kate Blanchfield}
 \affiliation{Department of Physics, Stockholm University, S-10691
 Stockholm, Sweden}
\author{Mohamed Bourennane}
 \affiliation{Department of Physics, Stockholm University, S-10691
 Stockholm, Sweden}
\author{Ingemar Bengtsson}
 \affiliation{Department of Physics, Stockholm University, S-10691
 Stockholm, Sweden}


\date{\today}



\begin{abstract}
Recent experiments have demonstrated ququart state-independent quantum contextuality and qutrit state-dependent quantum contextuality. So far, the most basic form of quantum contextuality pointed out by Kochen and Specker, and Bell, has eluded experimental confirmation. Here we present an experimentally feasible test to observe qutrit state-independent quantum contextuality using single photons in a three-path setup. In addition, we show that if the same measurements are performed on two entangled qutrits, rather than sequentially on the same qutrit, then the noncontextual inequality becomes a Bell inequality. We show that this connection also applies to other recently introduced noncontextual inequalities.
\end{abstract}



\maketitle


\section{Introduction}


Measurement outcomes of quantum $d$-level (with $d \ge 3$) systems prepared in an arbitrary state are not independent of compatible measurements performed on the same system \cite{Specker60,Bell66,KS67}. This is called quantum contextuality and can be experimentally observed through the violation of inequalities among probabilities of outcomes of compatible measurements satisfied by any noncontextual model (noncontextual inequalities). Recent experiments have shown state-independent quantum contextuality for quantum four-level systems \cite{KZGKGCBR09,ARBC09,MRCL09} following Ref.~\cite{Cabello08}. The simplest physical system in which quantum contextuality occurs is a quantum three-level system (qutrit). Quantum contextuality using a specific qutrit state has been recently observed \cite{LLSLRWZ11,AABC11,Cabello11} following Ref.~\cite{KCBS08}. However, the most basic form of state-independent quantum contextuality pointed out by Kochen and Specker (KS) \cite{Specker60,KS67} and Bell \cite{Bell66} has so far eluded experimental confirmation. The reason is that for all state-independent qutrit violations to date \cite{BBCP09,YO11,BBC11}, the violation is small and will be hidden by experimental imperfections, since its observation requires testing a large number of contexts.

In the first part of this paper we describe a specific experiment to observe qutrit state-independent contextuality with current technology.

A frequently raised criticism of experiments with sequential measurements focuses on the assumption of compatibility of the measurements. It adopts two forms: (i) that there is no operational definition of compatibility, and (ii) that there is no experimental way to guarantee that the sequential measurements are perfectly compatible---the so-called compatibility loophole \cite{GKCLKZGR10}. However, there is an operational definition of compatibility \cite{Peres95}; the problem lies in experimentally testing it, since it is difficult to implement the sharp repeatable quantum measurements assumed in the textbooks. One way to avoid the compatibility loophole is by performing local measurements on spatially-separated systems instead of sequential measurements on a single system, and convert noncontextual inequalities into Bell inequalities. This raises the question of the connection between contextuality and nonlocality. It is known that it is possible to convert some contextuality proofs based on KS sets into Bell inequalities \cite{Cabello01,Cabello10,CT11,AGACVMC11}. However, so far only one of them has a violation large enough to allow experimental verification with entangled ququarts \cite{AGACVMC11,CBPMD05,YZZYZZCP05}.

In the second part of this paper we derive a Bell inequality from a state-independent noncontextual inequality that does not contain a KS set. We derive an experimentally testable two-qutrit Bell inequality from state-independent quantum contextuality and obtain a Bell inequality from a state-independent quantum contextuality proof that does not contain a KS set. In addition, we show that the same method can be used to obtain Bell inequalities starting from any of the noncontextual inequalities recently introduced in Refs.~\cite{YO11,BBC11,YO12}.


\section{Experimentally testable state-independent noncontextual inequality}


Consider qutrit observables of the type
\begin{equation}
 \label{A}
 A_i=\openone - 2 |v_i \rangle \langle v_i |,
\end{equation}
where $\openone$ is the $3 \times 3$ identity matrix and $|v_i\rangle $ are unit rays. Each $A_i$ has the spectrum $\{ -1, 1, 1 \}$.
Following Yu and Oh \cite{YO11}, we choose
\begin{subequations}
 \label{V}
\begin{align}
 &|v_1\rangle=\frac{1}{\sqrt{3}} \begin{pmatrix}
      -1\\1\\1
    \end{pmatrix},\;
 |v_{5,6}\rangle=\frac{1}{\sqrt{2}}
  \begin{pmatrix}
      0\\1\\ \pm 1
    \end{pmatrix},\;
 |v_{11}\rangle=
  \begin{pmatrix}
      1\\0\\0
    \end{pmatrix},\\
  &|v_2\rangle=\frac{1}{\sqrt{3}} \begin{pmatrix}
      1\\-1\\1
    \end{pmatrix},\;
 |v_{7,8}\rangle=\frac{1}{\sqrt{2}}
  \begin{pmatrix}
      1\\0\\ \pm 1
    \end{pmatrix},\;
 |v_{12}\rangle=
  \begin{pmatrix}
      0\\1\\0
    \end{pmatrix},\\
 &|v_3\rangle=\frac{1}{\sqrt{3}}
  \begin{pmatrix}
      1\\1\\-1
    \end{pmatrix},\;
 |v_{9,10}\rangle=\frac{1}{\sqrt{2}}
  \begin{pmatrix}
      1\\ \pm 1\\0
    \end{pmatrix},\;
 |v_{13}\rangle=
  \begin{pmatrix}
      0\\0\\1
    \end{pmatrix},\\
 &|v_4\rangle =\frac{1}{\sqrt{3}}
 \begin{pmatrix}
      1\\1\\1
    \end{pmatrix}.
 \end{align}
 \end{subequations}
These 13 rays are a subset of the 33-vector set discovered by Peres \cite{Peres91} (and in fact span one of the three intersecting cubes appearing in a print by Escher \cite{Penrose00}).

The following inequality follows from the assumption that the outcomes of the measurements of $A_i$ are noncontextual values $1$ or $-1$,
\begin{equation}
 \label{ineq}
\begin{split}
 \kappa \equiv
 &\frac{1}{2} \left(\sum_{i=1}^4 \langle A_i \rangle - \sum_{i=1}^4 \sum_{j=5}^{10} \Gamma_{ij} \langle A_i A_j \rangle\right)\\
 &+\sum_{k=5}^{13} \langle A_k \rangle - \sum_{m=5}^{12} \sum_{n > m}^{13} \Gamma_{mn} \langle A_m A_n \rangle
 \le 9,
 \end{split}
\end{equation}
where $\Gamma_{ij}$ is 1 if $\langle v_i | v_j \rangle=0$, and 0 otherwise, and $\langle A_i A_j \rangle$ denotes the mean value of the product of the measurement outcomes. The upper bound has been verified by checking all possible assignments of values $1$ or $-1$.

Inequality \eqref{ineq} is a state-independent noncontextual inequality, since the prediction of quantum mechanics for any qutrit state (including the maximally mixed state $\rho = \frac{1}{3} \openone$) is
\begin{equation}
 \kappa_{\rm QM}=\frac{29}{3}=9+\frac{2}{3}.
\end{equation}
The form of the inequality is similar to that proposed by Yu and Oh \cite{YO11}, but has been improved for experimental implementation. Firstly, the relative weights of the terms in \eqref{ineq} improve the quantum violation. Secondly, the number of terms of the form $\langle A_i A_j \rangle$ has been reduced where possible for ease of measurement. Finally, the order of sequential measurements (where $A_i$ is measured first in the expression $A_i A_j$) minimizes the level of experimental complexity required for the qutrit system described below.

A state-independent noncontextual inequality is more robust against noise in the state preparation than a state-dependent one. This is because the violation is the same for any state, including the maximally mixed one. A good estimation of how feasible the inequality is for an actual experiment is the robustness of the violation against experimental imperfections. Assuming that all terms are similarly sensitive to errors, a reasonable measure of robustness is the difference between the quantum violation and the noncontextual bound divided by the number of terms scaled with their respective weights. With this definition, we calculate the robustness to noise of inequality \eqref{ineq} as
\begin{equation}
\frac{\frac{2}{3}}{\frac{1}{2}(4+12)+(9+12)} = \frac{2}{87} \approx 0.023 .
\end{equation}
The corresponding calculation for the inequality in the form first proposed by Yu and Oh \cite{YO11} yields $\frac{1}{75} \approx 0.013$.
The greater robustness to noise of inequality \eqref{ineq} shows that it is more suitable for an experimental test than any previous inequality \cite{BBCP09,BBC11}.


\section{Experimental realization of the qutrit state-independent contextuality test}

\begin{figure}[t]
\centerline{\includegraphics[width=4.8cm]{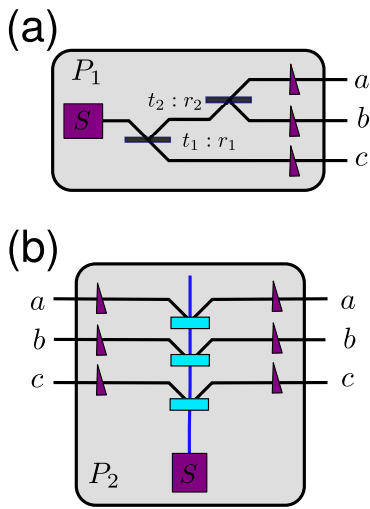}}
\caption{\label{Fig1}(Color online) (a) Set up for preparing arbitrary qutrit states ($P_1$). (b) Set up for preparing state \eqref{state} ($P_2$).}
\end{figure}
\begin{figure}[b]
\centerline{\includegraphics[width=8.0cm]{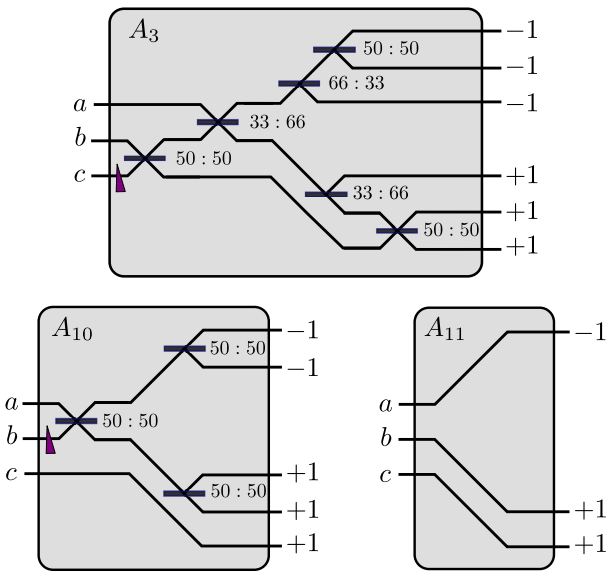}}
\caption{\label{Fig2}(Color online) Examples of the three classes of measurements needed for the experiment: $A_{3}$, $A_{10}$, and $A_{11}$, defined in \eqref{A} and \eqref{V}.}
\end{figure}

Testing inequality \eqref{ineq} is particulary simple on a qutrit defined by a single photon in a three-path ($a$, $b$, and $c$) setup. The basis vectors $|0\rangle, |1\rangle, |2\rangle $ correspond to finding the photon in path $a$, $b$, or $c$, respectively. Any state can be prepared with the setup in Fig.~\ref{Fig1}(a) by adjusting the transmittance and reflectance of the beam splitters (BSs) and tuning the wedges, which are related to the amplitude weights and the phase relations, respectively, of the state. The measurements $A_i$ can be implemented by mapping $|v_i\rangle$, the state corresponding to eigenvalue $-1$, to path $a$. This procedure will then map a state corresponding to eigenvalue $+1$ to the remaining two paths $b$ and $c$. When the transformation is implemented it is easy to separate $a$ from $b$ and $c$ for further processing. There are three types of measurement: (i) when $i=1,2,3,4$, (ii) when $i=5,\ldots,10$, and (iii) when $i=11,12,13$. The difference between two measurements of the same type is simply a path relabeling or a phase added, where the latter can be adjusted by a wedge. In Fig.~\ref{Fig2} we show an example of each type. To measure $A_i$ for $i = 5,\ldots,10$ a $50$:$50$ BS is used to transform $|v_i\rangle$ to the state $|0\rangle$ that corresponds to path $a$. The same principle holds for $i = 1,\ldots,4$, where the transformation to the state $|0\rangle$ is realized in two steps. First, a $50$:$50$ BS is used to transform the state to a superposition between $|0\rangle$ and $|1\rangle$. This is followed by a $33$:$66$ BS, which finishes the mapping to $|0\rangle$. The last type of measurement, $A_i$ for $i = 11, 12, 13$, is simple since it is already in the encoding basis, thus no transformation is needed except for rerouting the path to obtain $|v_i\rangle$ in path $a$. Before exiting each measurement box, a remapping is performed to keep the same encoding convention between the sequential measurements.


\begin{figure}[t]
\centerline{\includegraphics[width=8.0cm]{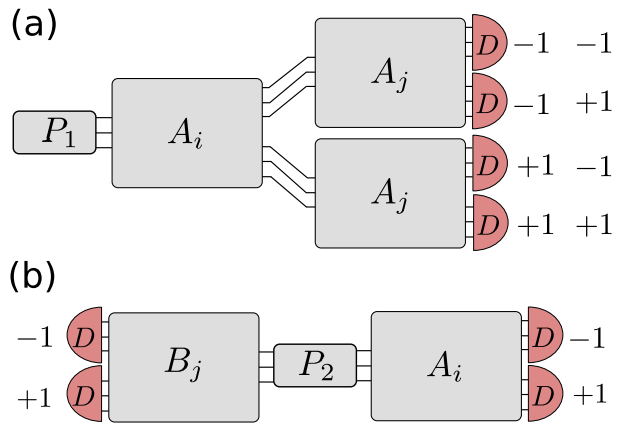}}
\caption{\label{Fig3}(Color online) (a) Cascade setup for sequentially measuring $A_i$ and $A_j$ to test inequality \eqref{ineq}. (b) Two-party set up for measuring $A_i$ and $B_j \equiv A_j$ on two different systems to test inequality \eqref{ineq3}.}
\end{figure}


Measuring $A_i A_j$ requires two sequential measurements on the same photon, first $A_i$ and then $A_j$. For that, we use a cascade setup illustrated in Fig.~\ref{Fig3}(a), in which the two outcomes of $A_i$ are each directed into identical but separated devices for measuring $A_j$. Each of the four possible combinations of outcomes corresponds to a detection in one of the four detectors. This approach satisfies the requirements for a noncontextuality test: (i) $A_i$ is measured using a similarly constructed device in every experiment (context), and (ii) the setup permits all possible combinations of results. For instance, while quantum mechanics predicts that in an ideal experiment the outcomes of two compatible $A_i$ and $A_j$ can never both equal 1, the setup allows such an event.

Like any test of Bell and noncontextual inequalities, the conclusions of the experiment may be affected by the detection loophole \cite{Pearle70}, unless the overall detection efficiency, defined as the ratio between the number of photons detected divided by the number of photons emitted by the source, is above a certain threshold, which depends on the inequality. This is in fact a difficult loophole to avoid with photons. The aim of this paper is only to show that qutrit state-independent contextuality can be observed with photons under the same assumptions made in most experiments on Bell inequalities, including the fair-sampling assumption. The proposal is not intended to be free of the detection loophole. A loophole-free version of the experiment will require additional features such as a preparation setup in which one can count the number of photons emitted and a good enough photodetection efficiency.


\section{Experimentally testable two-qutrit Bell inequality}


An interesting property of inequality \eqref{ineq} is that it is state-independent. Thus it is violated by any state, including any mixed state obtained by tracing out one qutrit from a two-qutrit entangled system. A second interesting property is that the quantum mechanical predictions for $\langle A_i A_j \rangle$ when the compatible observables $A_i$ and $A_j$ are measured sequentially on a single qutrit are the same as the predictions when $A_i$ and $A_j$ are measured on two separated qutrits prepared in the maximally entangled state
\begin{equation}
|\psi\rangle=\frac{1}{\sqrt 3}\left(|0\rangle_1 |0\rangle_2+|1\rangle_1 |1\rangle_2+|2\rangle_1 |2\rangle_2\right).
 \label{state}
\end{equation}
These two properties allow us to transform the noncontextual inequality \eqref{ineq} into a Bell inequality between two observers, Alice and Bob. The method has two steps. In the first step, the 13 observables are distributed between Alice and Bob: Alice measures $A_i$ with $i=1,\ldots,4,11,12,13$ and Bob measures $B_j \equiv A_j$ with $j=5,\ldots,10$. So inequality \eqref{ineq} becomes the noncontextual inequality
\begin{equation}
 \label{ineq2}
\begin{split}
 \kappa' \equiv
 &\frac{1}{2} \left(\sum_{i=1}^4 \langle A_i \rangle - \sum_{i=1}^4 \sum_{j=5}^{10} \Gamma_{ij} \langle A_i B_j \rangle\right)\\
 &+\sum_{j=5}^{10} \langle B_j \rangle +\sum_{k=11}^{13} \langle A_k \rangle - \sum_{k=11}^{13} \sum_{j=5}^{10} \Gamma_{kj} \langle A_k B_j \rangle\\
 &-\langle A_{11} A_{12} \rangle -\langle A_{11} A_{13} \rangle -\langle A_{12} A_{13} \rangle\\
 &-\langle B_{5} B_{6} \rangle-\langle B_{7} B_{8} \rangle-\langle B_{9} B_{10} \rangle
 \le 9,
 \end{split}
\end{equation}
which still contains terms involving two sequential measurements on Alice's qutrit (e.g., $\langle A_{11} A_{12} \rangle$) and on Bob's qutrit (e.g., $\langle B_{5} B_{6} \rangle$). In the second step, we exploit that, for state \eqref{state}, $\langle A_i B_i \rangle = 1$ for $i=1,\ldots,13$. 
By replacing both $\langle A_{i} A_{j} \rangle$ and $\langle B_{i} B_{j} \rangle$ with
$\frac{1}{2} \left(\langle A_{i} B_{j} \rangle + \langle A_{j} B_{i} \rangle - \langle A_{i} B_{i} \rangle - \langle A_{j} B_{j} \rangle\right)$,
we obtain the following Bell inequality:
\begin{equation}
 \label{ineq3}
\begin{split}
 \beta \equiv
 &\frac{1}{2} \left(\sum_{i=1}^4 \langle A_i \rangle - \sum_{i=1}^4 \sum_{j=5}^{10} \Gamma_{ij} \langle A_i B_j \rangle+\sum_{j=5}^{10} \langle A_{j} B_{j} \rangle\right.\\
 &\left.- \sum_{j=5}^{10} \sum_{m=5}^{10} \Gamma_{jm} \langle A_j B_m \rangle - \sum_{k=11}^{13} \sum_{n=11}^{13} \Gamma_{kn} \langle A_k B_n \rangle\right)\\
 &+\sum_{j=5}^{10} \langle B_j \rangle +\sum_{k=11}^{13} \langle A_k \rangle - \sum_{k=11}^{13} \sum_{j=5}^{10} \Gamma_{kj} \langle A_k B_j \rangle\\
 &+\sum_{k=11}^{13} \langle A_{k} B_{k} \rangle \le 15.
 \end{split}
\end{equation}
The upper bound follows from the locality assumption, i.e. the outcomes of the local measurements of $A_i$ and $B_j$ are independent of the measurement on the remote qutrit. Notice that to test inequality \eqref{ineq3}, Alice has to measure $A_i$ with $i=1,\ldots,13$, and Bob has to measure $B_j$ with $j=4,\ldots,13$. The prediction of quantum mechanics for state \eqref{state} is
\begin{equation}
 \beta_{\rm QM}=\frac{47}{3}=15+\frac{2}{3}.
\end{equation}

The Bell inequality \eqref{ineq3} is essentially different than any other two-qutrit Bell inequality previously tested \cite{TAZG04}: It connects the contextuality of a single qutrit to the nonlocality of an entangled pair. The measurements are the same in both tests (the only difference is the way they are performed---sequentially on the same system or on different systems) and the mean values are the same too [as $\langle A_{j} B_{j} \rangle=\langle A_{j} A_{j} \rangle$ for the state \eqref{state}]. The same two-step method for constructing the Bell inequality \eqref{ineq3} starting from the noncontextual inequality \eqref{ineq} can be applied to construct a Bell inequality starting from any of the recently introduced noncontextual inequalities in Refs.~\cite{YO11,BBC11,YO12}. The connection between the KS theorem and nonlocality shown in this paper differs from the one in Ref.~\cite{HR83} by not requiring the assumption of ideally perfect orthogonalities and state preparations, and thus, is experimentally testable. For example, nonlocality can be observed even if, instead of state $|\psi\rangle$ given by \eqref{state}, one prepares a noisy state $V|\psi\rangle \langle \psi|+(1-V) \frac{1}{9} \openone$ with $V > 0.95$, which is in the range of a carefully designed experiment.


\section{Experimental realization of the Bell test}


A setup for preparing the desired two-photon state \eqref{state} is shown in Fig.~\ref{Fig1}(b). A laser pumps three successive nonlinear crystals, each of which can spontaneously create a pair of photons with equal probability. A photon pair can be created in the first, second, or third crystal. Each term $|k\rangle_1 |k\rangle_2$ in state \eqref{state} is directly related to which crystal the photon pair was created in. The coherent superposition of state \eqref{state} is created by keeping stable the relative phases between the arms. The left- and right-hand sides of the preparation box are distributed to Alice and Bob respectively as shown in Fig.~\ref{Fig3}(b). Using the distributed state and the measurement setups described in Fig.~\ref{Fig2}, Alice and Bob can perform coincidence measurements to evaluate inequality \eqref{ineq3}.


\section{Conclusions}


We have introduced a state-independent noncontextual inequality for qutrits with a quantum violation large enough to be observed in a real experiment. This will allow an experimental observation of the most basic and fundamental form of quantum contextuality. In addition, we have shown that the same measurements used for the state-independent contextuality test on a single qutrit allow us to test nonlocality on a pair of entangled qutrits, and that this also applies to any of a family of recently introduced noncontextual inequalities. This establishes a connection between both types of experiments and provides examples of experimentally testable Bell inequalities constructed from a proof of qutrit state-independent contextuality, and examples of Bell inequalities constructed from proofs of state-independent contextuality that do not require KS sets.


\begin{acknowledgments}
This work was supported by the Swedish Research Council (VR), the Linnaeus Center of Excellence ADOPT, the Projects No.\ FIS2008-05596 and No.\ FIS2011-29400, and the Wenner-Gren Foundation.
\end{acknowledgments}




\begin{thebibliography}{99}


\bibitem{Specker60}
 E. P. Specker,
 Dialectica \textbf{14}, 239 (1960).

\bibitem{Bell66}
 J. S. Bell,
 Rev. Mod. Phys. \textbf{38}, 447 (1966).

\bibitem{KS67}
 S. Kochen and E. P. Specker,
 J. Math. Mech. \textbf{17}, 59 (1967).


\bibitem{KZGKGCBR09}
 G. Kirchmair, F. Z\"ahringer, R. Gerritsma, M. Kleinmann,
 O. G{\"u}hne, A. Cabello, R. Blatt, and C. Roos,
 Nature (London) \textbf{460}, 494 (2009).

\bibitem{ARBC09}
 E. Amselem, M. R{\aa }dmark, M. Bourennane, and A. Cabello,
 Phys. Rev. Lett. \textbf{103}, 160405 (2009).

\bibitem{MRCL09}
 O. Moussa, C. A. Ryan, D. G. Cory, and R. Laflamme,
 Phys. Rev. Lett. \textbf{104}, 160501 (2010).


\bibitem{Cabello08}
 A. Cabello,
 Phys. Rev. Lett. \textbf{101}, 210401 (2008).


\bibitem{LLSLRWZ11}
 R. {\L}apkiewicz, P. Li, C. Schaeff, N. Langford, S.~Ramelow, M. Wie\'{s}niak,
 and A. Zeilinger,
 Nature (London) \textbf{474}, 490 (2011).

\bibitem{AABC11}
 J. Ahrens,
 E. Amselem,
 M. Bourennane, and
 A. Cabello (unpublished).

\bibitem{Cabello11}
 A. Cabello,
 Nature (London) \textbf{474}, 456 (2011).


\bibitem{KCBS08}
 A. A. Klyachko, M. A. Can, S. Binicio\u{g}lu, and A. S.~Shumovsky,
 Phys. Rev. Lett. {\bf 101}, 020403 (2008).


\bibitem{BBCP09}
 P. Badzi{\c a}g, I. Bengtsson, A. Cabello, and I. Pitowsky,
 Phys. Rev. Lett. \textbf{103}, 050401 (2009).

\bibitem{YO11}
 S. Yu and C. H. Oh,
 Phys. Rev. Lett. \textbf{108}, 030402 (2012).

\bibitem{BBC11}
 I. Bengtsson, K. Blanchfield, and A. Cabello,
 Phys. Lett.~A \textbf{376}, 374 (2012).


\bibitem{GKCLKZGR10}
 O. G{\"u}hne, M. Kleinmann, A. Cabello, J.-\AA. Larsson,
 G.~Kirchmair, F. Z\"ahringer, R. Gerritsma, and C.~F.~Roos,
 Phys. Rev.~A \textbf{81}, 022121 (2010).

\bibitem{Peres95}
 A. Peres,
 \emph{Quantum Theory: Concepts and Methods}
 (Kluwer, Dordrecht, 1995), p. 203.

\bibitem{Cabello01}
 A. Cabello,
 Phys. Rev. Lett. \textbf{87}, 010403 (2001).

\bibitem{Cabello10}
 A. Cabello,
 Phys. Rev. Lett. \textbf{104}, 220401 (2010).

\bibitem{CT11}
 A. Cabello and M. Terra Cunha,
 Phys. Rev. Lett. \textbf{106}, 190401 (2011).

\bibitem{AGACVMC11}
 L. Aolita, R. Gallego, A. Ac\'{\i}n,
 A. Chiuri, G. Vallone, P.~Mataloni, and A. Cabello,
 Phys. Rev. A (in press) (2012); \eprint{arXiv:1105.3598}.

\bibitem{CBPMD05}
 C. Cinelli, M. Barbieri, R. Perris, P. Mataloni, and F. De Martini,
 Phys. Rev. Lett. \textbf{95}, 240405 (2005).

\bibitem{YZZYZZCP05}
 T. Yang, Q. Zhang, J. Zhang, J. Yin, Z. Zhao, M.~\.{Z}ukowski, Z.-B. Chen, and J.-W. Pan,
 Phys. Rev. Lett. \textbf{95}, 240406 (2005).

\bibitem{YO12}
 Inequalities (2) and (3) in S. Yu and C. H. Oh,
 \eprint{arXiv:1112.5513}.


\bibitem{Peres91}
 A. Peres,
 J. Phys. A \textbf{24}, L175 (1991).

\bibitem{Penrose00}
 R. Penrose,
 in \emph{Quantum Reflections},
 edited by J. Ellis and D. Amati
 (Cambridge University Press, Cambridge, 2000), p.~1.

\bibitem{Pearle70}
 P. M. Pearle,
 Phys. Rev. D \textbf{2}, 1418 (1970).

\bibitem{TAZG04}
 R. T. Thew, A. Ac\'{i}n, H. Zbinden, and N. Gisin,
 Phys. Rev. Lett. \textbf{93}, 010503 (2004).

\bibitem{HR83}
 P. Heywood and M. L. G. Redhead,
 Found. Phys. \textbf{13}, 481 (1983).


\end{thebibliography}
\end{document}